\documentclass[conference]{ieeeconf}
\IEEEoverridecommandlockouts
\usepackage{cite}
\usepackage{amsmath,amssymb,amsfonts}
\usepackage{algorithmic}
\usepackage{subcaption}
\usepackage{graphicx}
\usepackage{textcomp}
\usepackage{booktabs}
\usepackage{makecell}
\usepackage{xcolor}
\usepackage{url}

\usepackage{hyperref}

\hypersetup{hidelinks}

\def\BibTeX{{\rm B\kern-.05em{\sc i\kern-.025em b}\kern-.08em
    T\kern-.1667em\lower.7ex\hbox{E}\kern-.125emX}}
\begin{document}

\title{\LARGE\bf Model Predictive Control Design for Unlocking the Energy Flexibility of Heat Pump and Thermal Energy Storage Systems
}

\author{Weihong Tang$^1$, Yun Li$^1$, Shalika Walker$^2$, Tamas Keviczky$^1$
\thanks{$^1$W. Tang, Y. Li and T. Keviczky are with the Delft Center for Systems and Control, 
        Delft University of Technology,  Delft, the Netherlands.
        {\tt\small ttonytany@gmail.com; y.li-39@tudelft.nl; T.Keviczky@tudelft.nl}}
\thanks{$^2$S. Walker is with the Kropman BV, the Netherlands.
{\tt\small shalika.walker@kropman.nl}}
\thanks{This work was supported by the Brains4Buildings project under the Dutch grant programme for Mission-Driven Research, Development and Innovation (MOOI).}
}

\maketitle

\begin{abstract}
Heat pump and thermal energy storage (HPTES) systems, which are widely utilized in modern buildings for providing domestic hot water, contribute to a large share of household electricity consumption. With the increasing integration of renewable energy sources (RES) into modern power grids, demand-side management (DSM) becomes crucial for balancing power generation and consumption by adjusting end users' power consumption. This paper explores an energy flexible Model Predictive Control (MPC) design for a class of HPTES systems to facilitate demand-side management. The proposed DSM strategy comprises two key components: i) \textit{flexibility assessment}, and ii) \textit{flexibility exploitation}. Firstly, for \textit{flexibility assessment}, a tailored MPC formulation, supplemented by a set of auxiliary linear constraints, is developed to quantitatively assess the flexibility potential inherent in HPTES systems. Subsequently, in \textit{flexibility exploitation}, the energy flexibility is effectively harnessed in response to feasible demand response (DR) requests, which can be formulated as a standard mixed-integer MPC problem. Numerical experiments, based on a real-world HPTES installation, are conducted to demonstrate the efficacy of the proposed design. 
\end{abstract}

\section{Introduction}
Climate change has emerged as a pressing global concern, and prompts global actions. To achieve carbon neutrality and sustainable energy usage, the European Union has set an ambitious goal to increase the share of renewable energy sources (RES), such as solar energy and wind energy, to at least 27\% by 2030 \cite{dutchgov2022}. However, it is well-known that due to the intermittency and volatility of RES, growing reliance on the electricity generation from RES can introduce significant challenges in balancing supply and demand in the power grid, and hence lead to congestion. As a response, the concept of demand-side management (DSM) is proposed, which entails adjusting the energy consumption of end users according to the needs of the power grid, to balance the real-time power generation and consumption \cite{bunning2022robust,Abl:45}.

As an energy efficient and low-carbon thermal generation option, heat pumps (HP) are widely adopted in Europe to provide thermal energy in buildings, such as floor heating and domestic hot water usage, and the number of installations are on a sharp rise. It is estimated that by 2027, 10 million additional heat pumps will be installed in the EU \cite{hp2022}. A typical utilization of HPs is the combination of HP and thermal storage system, which we refer to as HPTES in this paper. With the support of TES, the thermal energy generated by HPs can be stored, which largely increases the flexibility of electricity consumption of HPs \cite{ermel2022thermal}. Considering the large amount of energy consumption of HPs in buildings and the energy flexibility emanating from HPTES systems, it is of interest to investigate potential improvements or replacement of the existing rule-based control schemes in HPTES systems for achieving DSM. As an advanced control technique, model predictive control (MPC) is regarded as a promising and effective control scheme to be widely implemented in future building mangement systems (BMS) due to its versatality in coping with system constraints, economic considerations, as well as the alignment of predicted energy consumption and availability. In \cite{alimohammadisagvand2016cost,tang2019model,kircher2015model,renaldi2017optimisation} and references therein, MPC-based operation schemes are designed for HPTES systems.

While MPC-based schemes for economic operation of HPTES have been extensively studied in literature and are shown to be effective, when considering DSM in the form of demand response (DR) requests, there is room for further improvement and alternative control schemes. The existing works about DSM of HPTES follow the so-called incentive-based or price-based programs \cite{d2019mapping,golmohamadi2022integration}. As pointed out in \cite{li2023unlocking}, while such schemes are relatively easy to implement, these programs fall short of full exploitation of energy flexibility and might fail to achieve the expected energy reduction from the power grid. In order to better utilize the energy flexibility of buildings and meet the demands of the power grid, a two-step DSM framework is proposed in \cite{li2023unlocking}. This scheme entails assessing the energy flexibility potential of the system, and a bidirectional communication between BMS and the power grid. Also, this two-step DSM framework fits in the S2 standard that is developed by the European Commission for exploiting energy flexibility in built environment \cite{S22023}.

In this paper, we will design an MPC-based DSM scheme based on the framework proposed in \cite{li2023unlocking} and investigate an energy flexible MPC design for a class of HPTES systems to achieve demand-side management. A general control-oriented model for the HPTES system is introduced. Based this model, a unified MPC framework for quantitatively assessing the energy flexibility potential of the HPTES system and exploting the flexibility via DR requests are designed. The proposed flexibility assessment approach only entails considering extra linear constraints within a typical economic MPC problem. With the proposed approach, the energy flexibility emanated from the thermal storage tanks is utilized to solve grid congestion problems without violating system constraints.

The remaining parts of this paper are organized as follows. Section \ref{sec:models} describes a control-oriented model for the HPTES system. Section \ref{sec:mpc} provides a unified MPC design framework for assessing and exploiting the energy flexibility of the HPTES system. Simulation results are presented in Section \ref{sec:simu}, followed by Conclusions in Section \ref{sec:conclusion}. 

\section{HPTES System Modelling and Problem Formulation}\label{sec:models}
\begin{figure}
\centering
    \includegraphics[width = \linewidth]{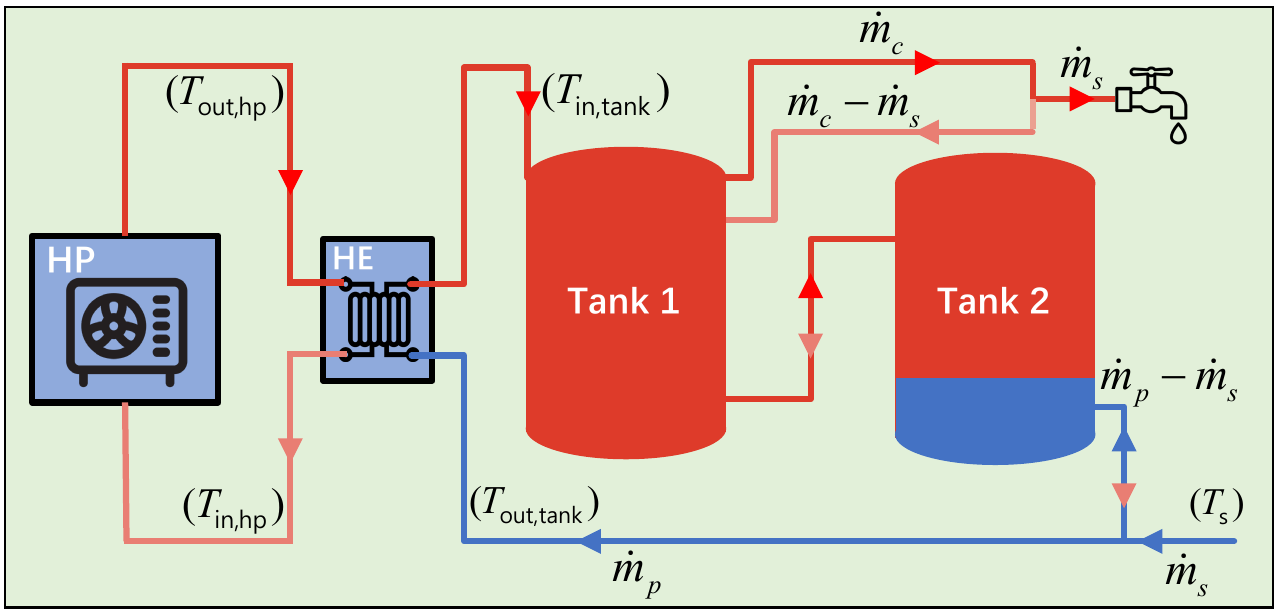}
    \caption{Diagram of the considered general HPTES system.}
    \label{fig:HPTES}
\end{figure}

In this section, a general control-oriented model of the HPTES system and the corresponding physical constraints are introduced. The HPTES system considered in our work is composed of three main components: a heat pump, a heat exchanger (HE), and two thermal storage tanks. The configuration of the HPTES system that will be investigated is shown in Fig. \ref{fig:HPTES}, where blue lines represent cold water flows, red lines represent hot water flows, and arrows represent the flow directions. The mass flow rates and temperature of corresponding pipelines that are used in developing the HPTES model are also indicated. This configuration is based on a  typical real HPTES system installation located in an office building in the Netherlands.

Hot water from the top of Tank 1 is circulated at a mass flow rate $\dot{m}_c$ going through connected taps to meet hot water usage demands. The real-time hot water usage is $\dot{m}_s$, whose future values are assumed to be predictable with sufficient accuracy from historical data. When hot water is consumed, the same amount of cold water $\dot{m}_s$ is supplied to the bottom of Tank 2  at the same time so that the total amount of water in the HPTES system is always constant. If the HP is off, the cold water will be directly injected into the bottom of Tank 2; otherwise, the water from the bottom of Tank 2 with mass flow rate $\dot{m}_p$ will be heated by HP then pumped into the top of Tank 1. The corresponding mass flow rates of the water circulations are indicated in Fig. \ref{fig:HPTES}.
 
In the following, we will provide a general control-oritened model of the above HPTES system, which will be used for our MPC design in Section \ref{sec:mpc}. 

\subsection{Heat Pump Model}
Heat pumps are used for providing thermal energy in the HPTES system. To model the relationship between the electricity consumption and thermal energy generation of HP, the concept of coefficient of performance (COP) is used. The COP is defined as the ratio of heat output to work input, which can be expressed as
\begin{equation}\label{eq:cop}
    \text{COP} = \frac{Q_{\text{hp}}}{P_{\text{hp}}},
\end{equation}
where $Q_{\text{hp}}$ denotes the heat output of the heat pump and $P_{\text{hp}}$ its power input. 

It is assumed in our design that the HP only operates in two modes: on and off. Hence, it follows from \eqref{eq:cop}, that the thermal energy generated by HP can be denoted as
\begin{equation}\label{eq:hp}
    Q_{\text{hp}} = \text{COP}\cdot P_{r}\cdot u,
\end{equation}
where $P_{r}$ is the rated power of the heat pump, and $u\in\mathbb{B}$ is the on/off binary control signal.

In order to balance approximation accuracy and computational burden of the resulting MPC problem, the following bilinear COP model is adopted:
\begin{equation}\label{eq:cop2}
\text{COP}=a_1+a_2 \cdot T_{\text{in,hp}}+a_3 \cdot T_{\text{amb}}+a_4 \cdot T_{\text{in,hp}} \cdot T_{\text{amb}},
\end{equation}
where $T_{\text{in,hp}}$ is the inlet water temperature of HP, $T_{\text{amb}}$ is the ambient air temperature of HP, and $(a_1,a_2,a_3,a_4)$ are parameters to be identified.

\subsection{Thermal Energy Storage Model}
\begin{figure}[h!]
            \centering
    \includegraphics[width=0.5\linewidth]{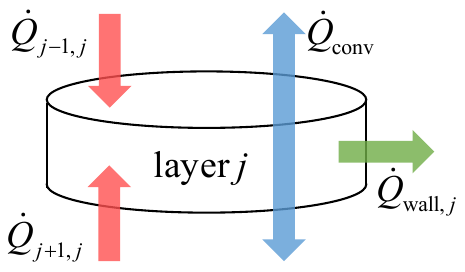}
            \caption{Heat flow diagram of a single layer of TES.}
            \label{fig:node}
\end{figure}

The thermal energy generated by HP is used to heat the water in the connected water tanks. For modeling the water tanks, the stratified tank model is adopted to balance modeling accuracy and computational burden \cite{rastegarpour2018predictive}. In the stratified water tank model, a single water tank is divided into several connected layers, or called nodes. Each layer is assumed to function as an individual node with uniform temperature distribution. 
% In our study, we employed the Stratified Tank Model to represent our water storage tanks. This model envisions the tank as divided into $N$ discrete, equivalent nodes or layers, with each layer functioning as an individual node characterized by a uniform temperature. The stratification method is notably beneficial for precisely depicting the thermal dynamics within the tank, effectively balancing model complexity with accuracy.
The general thermal behavior of each layer is depicted in Fig. \ref{fig:node}, where the thermal exchange of each node is composed by the following parts:
\begin{itemize}
    \item $\dot{Q}_{\text{wall},j}$: the heat loss from layer $j$ to the tank wall.
    \item $\dot{Q}_{j+1,j}$ and $\dot{Q}_{j-1,j}$: thermal conduction between layer $j$ and its adjacent layers $j+1$ and $j-1$.    
    \item $\dot{Q}_{\text{conv}}$: heat exchange due to water flows $\dot{m}_t$ in/out of $j$-th layer from/to adjacent layers.
\end{itemize}

Then, the thermal dynamics of each thermal node can be written as
\begin{subequations}\label{eq:tank_model}
\begin{align}
& m_{j} c_p \frac{dT_j}{dt}=-\dot{Q}_{\text {wall}, j}+\dot{Q}_{j+1,j}+\dot{Q}_{j-1,j}+\dot{Q}_{\text {conv}}, \\
& \dot{Q}_{\text {conv}}\!=\!
\begin{cases}\dot{m}_t c_{p}(T_{j-1}-T_{j}),\ \dot{m}_t\text{ flows from $j\!-\!1$ to $j$ } \\
\dot{m}_tc_p(T_{j+1} -T_j),\ \dot{m}_t\text{ flows from $j\!+\!1$ to $j$}
\end{cases}\\
& \dot{Q}_{j+1,j}=R_{j,j+1}\left(T_{j+1}-T_{j}\right), \\
& \dot{Q}_{j-1,j}=R_{j-1,j}\left(T_{j-1}-T_{j}\right) , \\
& \dot{Q}_{\text {wall},j}=R_w\left(T_{j}-T_{\text{amb}}\right),
&
\end{align}
\end{subequations}
where $m_j$ is the mass of the $j$-th node, $c_p$ is the specific heat capacity of water, $\dot{m}_t$ is the water mass flow rate between adjacent tank layers, $R_{j,j+1}$ and $R_w$ are thermal resistances between the $j$-th layer and the $(j+1)$-st layer and the tank wall, respectively.

Equation \eqref{eq:tank_model} describes the general thermal behaviour of a water layer. For layers with hot water extraction and cold water supplement, their thermal dynamics need extra modification. For the top layer of Tank 1, it does not have an upper layer so the thermal conduction $\dot{Q}_{j-1,j} = 0$, and the heat convection $\dot{Q}_{\text{conv}} = \dot{m}_tc_p(T_{j-1}-T_j)$ should be computed with $T_{j-1}$ as the temperature of the inlet water injected into Tank 1. The same analysis is also applicable to the bottom layer of Tank 2, where no lower layer exists. In addition, in our system, since the water circulation direction will change with the operation of HP, namely $u$, the mass flow rate of convection, $\dot{m}_t$, is a function of HP control signal $u$ and the mass flow rate of hot water consumption $\dot{m}_s$.  

% In addition to the general thermal behavior described, it's crucial to account for the dynamics of mass flow within the tanks, which vary based on the operational status of the heat pump. When the heat pump is active, water is circulated from Tank 1 to Tank 2 at a rate of $\dot{m}_p-\dot{m}_s$, indicating the net flow considering the heat pump's contribution and any simultaneous draw from the cold water supply pipe. Conversely, when the heat pump is inactive, replenishment occurs solely through the cold water supply, with water flowing from Tank 2 back to Tank 1 at a rate of $\dot{m}_s$, representing the supply draw.

% Furthermore, specific layers within the tanks, the top layer of Tank 1 and the bottom layer of Tank 2, has unique thermal conduction characteristics. For the top layer of Tank 1, thermal conduction is limited to $Q_{j,j+1}$, with $T_{j-1}$ set to the inlet water temperature of the water tanks. Similarly, in the bottom layer of Tank 2, thermal conduction is solely represented by $Q_{j-1,j}$, and $T_{j+1}$ corresponds to the temperature of the supply pipe. 

% {\color{red}describe difference of mass flow when heat pump is on and off (heat pump on, from tank 1 to tank2 with $\dot{m}_p-\dot{m}_s$; heat pump is off, from tank 2 to tank 1 with $\dot{m}_s$), and the case for the tank 1 top and tank 2 bottom.}

The thermal dynamics in \eqref{eq:tank_model} is able of capturing the thermal behaviour of a general thermal storage system and can be easily adapted to other water tank configurations with minor modification. Also, this model is sufficiently simple to be incorporated into MPC design as a prediction model.

% This energy balance equation is instrumental in capturing the dynamic thermal behavior and interactions within and across the layers, offering a simplified yet precise depiction of the thermal dynamics within the tanks. Its simplicity ensures ease of integration into broader system simulations, maintaining a desirable level of accuracy in reflecting the tanks' thermal behavior.

% \subsection{Heat Exchanger}

\subsection{System Constraints}
For safe operation and demand requirements satisfaction, the HPTES system is subject to several physical constraints. In this subsection, we describe the system constraints that the proposed control scheme has to satisfy. 

Firstly, to provide qualified hot water and ensuring safe operation, the water temperature of the HPTES system has to be within an admissible range, which is denoted as
\begin{equation}\label{eq:temp_cons}
    x^l \leq x \leq x^u,
\end{equation}
where $x$ denotes the temperature vector of different water layers and pipelines, $x^l$ and $x^u$ are the corresponding temperature lower bound and upper bound, respectively.

Secondly, frequent on-off switches of HP is unfavourable since it can cause excessive wear and tear and reduce the HP lifetime. Consequently, the proposed MPC scheme should prevent frequently switching the HP. For this purpose, the following constraint is added:
\begin{equation}\label{eq:switch}
\sum_{k=1}^{m}(u_{t-k+1}-u_{t-k})^2 \leq N_{\text {ctrl}},
\end{equation}
where $N_{\text{ctrl}}$ is the maximal number of switches that is allowed during $m$ time steps, and $u_{t-k}$ denotes the control input $k$ time steps in the past compared to the current time instant $t$. Because $u_t\in\mathbb{B}$, the LHS of \eqref{eq:switch} computes the number of switches during the most recent $m$ time steps. For instance, setting $m=8$ and $N_{\text {ctrl}}=1$ implies that, within an 8-step time window, only one switch is allowed.

\textbf{Remark 1}: The HPTES configuration and the corresponding control-oriented models and system constraints described in this section are general enough to cover a wide range of possible HPTES systems. Also, the control-oriented models can be flexibly adjusted while still keeping a similar model structure even if other alternative configurations are adopted. For example, the position of hot water extraction, which is at the top of Tank 1 in our case, and the position of cold water supplement, which is at the bottom of Tank 2 in our case, can be placed in other positions. Consequently, our MPC-based control design scheme in Section \ref{sec:mpc} will also be applicable to these alternative configurations with only minor modifications.

\section{Energy Flexible MPC Design for Demand-Side Management}\label{sec:mpc}
\begin{figure}
    \centering
    \includegraphics[width=0.85\linewidth]{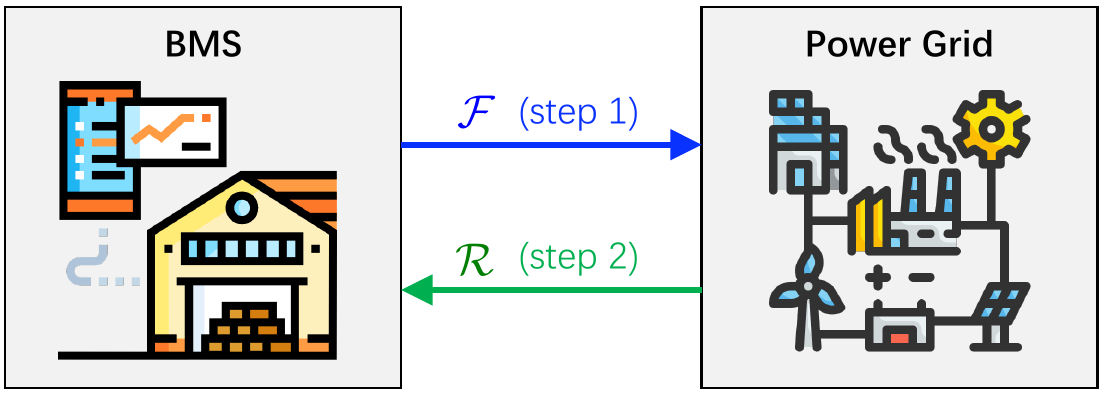}
    \caption{Diagram of the two-step DSM scheme.}
    \label{fig:DSM}
\end{figure}

In this section, an MPC strategy is designed for the HPTES system by following the two-step DSM framework shown in Fig. \ref{fig:DSM}. The DSM scheme consists of two steps: flexibility assessment and flexibility exploitation. 

As pointed out in \cite{d2019mapping,golmohamadi2022integration,li2023unlocking}, most of the DSM strategies follow either price-based or incentive-based programs, which are both one-step approaches and fail to fully exploit the energy flexibility of the system. 
With this two-step DSM framework, at \textit{step 1}, the flexibility potential of the HPTES system is assessed by BMS and the flexibility information $\mathcal{F}$ is sent to the grid operator. Then, at \textit{step 2}, the grid operator generates a feasible DR request $\mathcal{R}$ to the BMS to activate the energy flexibility. In contrast to the existing price-based or incentive-based programs, this two-step DSM approach guarantees that the expected DR requests from the power grid will be achieved. This is achieved by quantitatively assessing the flexibility potential of the system before exploitation, and generating feasible DR requests. This property is beneficial in efficiently coordinating different DR service providers and conducting DSM.
\subsection{Flexibility Assessment}
The first question to answer for the two-step DSM scheme is how to assess the energy flexibility potential of the HPTES system that is performed in step 1. 
It has been proposed that there are three basic factors in determining energy flexibility: magnitude, time duration, and cost \cite{junker2018characterizing,soren17}. Consequently, we will quantitatively describe the energy flexibility of the HPTES system via these factors.

Since we focus on solving the grid congestion problem, which is currently a challenging issue in the Dutch electricity market, the exploitation of flexibility refers to reducing the electricity consumption of HP. For our considered HPTES system, the control input of HP only has two modes: on and off. Consequently, the capacity of flexibility is proportional to the time duration that HP can remain off. Accordingly, in our work, the goal of flexibility assessment is to determine the optimal flexibility duration to maximize the benefits of providing DR services, i.e., the period during which HP is off without violating system constraints. 

\begin{figure}
    \centering
    \includegraphics[width = 0.8\linewidth]{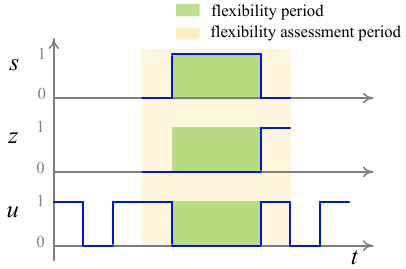}
    \caption{Schematic of the logic in \eqref{eq:flexi_cons} indicating flexibility periods.}
    \label{fig:logics}
\end{figure}

To conduct flexibility assessment, we introduce two sets of auxiliary decision variables $\mathcal{S}_{\mathcal{T}}:= \{s_t,t\in\mathcal{T}\}$ and $\mathcal{Z}_{\mathcal{T}}:= \{z_t,t\in\mathcal{T}\}$, where $s_t\in\mathbb{B}$ and $z_t\in\mathbb{B}$ are binary variables, and $\mathcal{T}:=\{t_{f_1},\cdots,t_{f_n}\}$ is the set of time indices during which the flexibility is assessed, called flexibility assessment period. Based on $s_t$ and $z_t$, we design the following logic:
\begin{itemize}
    \item $s_t = 1$ and $z_t = 0$: time instant $t$ is within the flexibility period, and hence $u_t = 0$.
    \item $s_t = 0$ and $z_t = 0$: time instant $t$ is before the flexibility period, and $u_t\in\mathbb{B}$.
    \item $s_t = 0$ and $z_t = 1$: time instant $t$ is after flexibility period, and $u_t\in\mathbb{B}$.
\end{itemize}
The above logic can be mathematically formulated as the following linear constraints:
\begin{subequations}\label{eq:flexi_cons}
    \begin{align}
        &u_t \leq 1 - s_t, \\
        &s_{t+1} \geq s_t - z_{t+1}, \\
        &s_t + z_t \leq 1,\\
        &z_{t+1} \geq z_t.
    \end{align}
\end{subequations}
In Fig. \ref{fig:logics}, a schematic illustration of the above logics is provided, where the flexibility assessment period is defined by $\mathcal{T}$, and the flexibility period is defined by $\mathcal{F}$, whose formal definition is given below.
\begin{equation}\label{eq:flexi_period}
\mathcal{F}:=\{t|s_t = 1, t \in\mathcal{T}\}\subseteq \mathcal{T}.
\end{equation}
The flexibility period $\mathcal{F}$ gives the optimal flexibility capacity, i.e., the time period within $\mathcal{T}$ that the HP can be off without violating system constraints, that the BMS will promise to the grid operator.

Finally, flexibility assessment can be mathematically formulated as the following optimization problem
\begin{subequations}\label{eq:flexi_comp}
    \begin{align}
       \min_{ u_t,s_k,z_k}\ & J_o - J_f \\
        \text{s.t. } & x_{t+1} = f(x_{t},u_{t},d_{t}), \label{eq:sys_dyn}\\
        & x_t \text{ satisfy } \eqref{eq:temp_cons},\ u_t \text{ satisfy }\eqref{eq:switch},\\
        & \forall t = 0,1,2,\cdots, N-1,\\
        & (u_k,s_k,z_k)\text{ satisfy }\eqref{eq:flexi_cons}, \forall k\in\mathcal{T}
    \end{align}
\end{subequations}
where $N$ is the length of prediction horizon, $J_o = \sum_{t= 1}^N l_t(x_t,u_{t-1})$ is the total operational cost of the HPTES system within the prediction horizon with $l_t(x_t,u_{t-1})$ as the $t$-th stage cost, $J_f = \lambda\sum_{t\in\mathcal{T}} s_t$ is the revenue of flexibility that is proportional to the length of flexibility period, $\lambda\in\mathbb{R}_+$ is a scaling factor, $f(x_t,u_t,d_t)$ is the control-oriented model of the HPTES system introduced in \eqref{eq:hp} and \eqref{eq:tank_model} with system states $x_t$ as the temperature vector of different water layers and pipelines, $d_t$ as the mass flow rate of hot water consumption $\dot{m}_s$ at time instant $t$, and $u_t$ the binary control input of HP. For the sake of brevity, the detailed expressions of $f(x_t,u_t,d_t)$ are given in Section \ref{sec:appx}.
 
% If the operational cost $J_o$ is omitted in \eqref{eq:flexi_comp}, then the corresponding flexibility period $\mathcal{F}$ is the longest time period that the HP can remain off with no constraint violation. 

\textbf{Remark 2}: The optimization problem \eqref{eq:flexi_comp} aims at optimally balancing the operational cost and the benefit of promising a certain capacity for flexibility as DSM to the power grid. If the operational cost term $J_o$ in the objective function is omitted, solving \eqref{eq:flexi_comp} will allow us to calculate the maximal capacity of energy flexibility. Namely, the longest consecutive time period during which the HP can remain off without violating system constraints. It should be noted that, compared with a typical economic MPC problem formulation, the proposed flexibility assessment problem only entails introducing extra linear constraints \eqref{eq:flexi_cons}, and is independent of the structure of the system dynamics \eqref{eq:sys_dyn}. These properties are beneficial to the scalability and applicability of the proposed scheme.

\subsection{Flexibility Exploitation}
After solving the optimization problem \eqref{eq:flexi_comp}, BMS will send the promised flexibility capacity $\mathcal{F}$ to the power grid. Based on this information, the grid operator will subsequently generate a feasible DR request $\mathcal{R} = \{t|t\in\mathcal{F}\}\subseteq\mathcal{F}$ to specify the exact time instants that the HP is required to be off. Namely,
\begin{equation}\label{eq:dr_request}
    u_t = 0,\forall t\in\mathcal{R}.
\end{equation}
For achieving this DR request, in BMS a general MPC problem can be formulated by imposing the constraints \eqref{eq:dr_request}. An example for such a formulation will be shown in \eqref{eq:dr_comp} of Section \ref{sec:simu}. Since the DR request $\mathcal{R}$ is feasible w.r.t. the flexibility capacity $\mathcal{F}$, i.e., $\mathcal{R}\subseteq \mathcal{F}$, it can be guaranteed that the DR request in \eqref{eq:dr_request} can be achieved without violating system constraints owing to the fact that $u_t = 0$ $(\forall t \in\mathcal{F})$ is always a feasible solution. 

\section{Simulation Results}\label{sec:simu}
\begin{figure}[!h]
    \centering
\begin{subfigure}[b]{0.45\linewidth}
    \centering
        \includegraphics[height=0.7\linewidth]{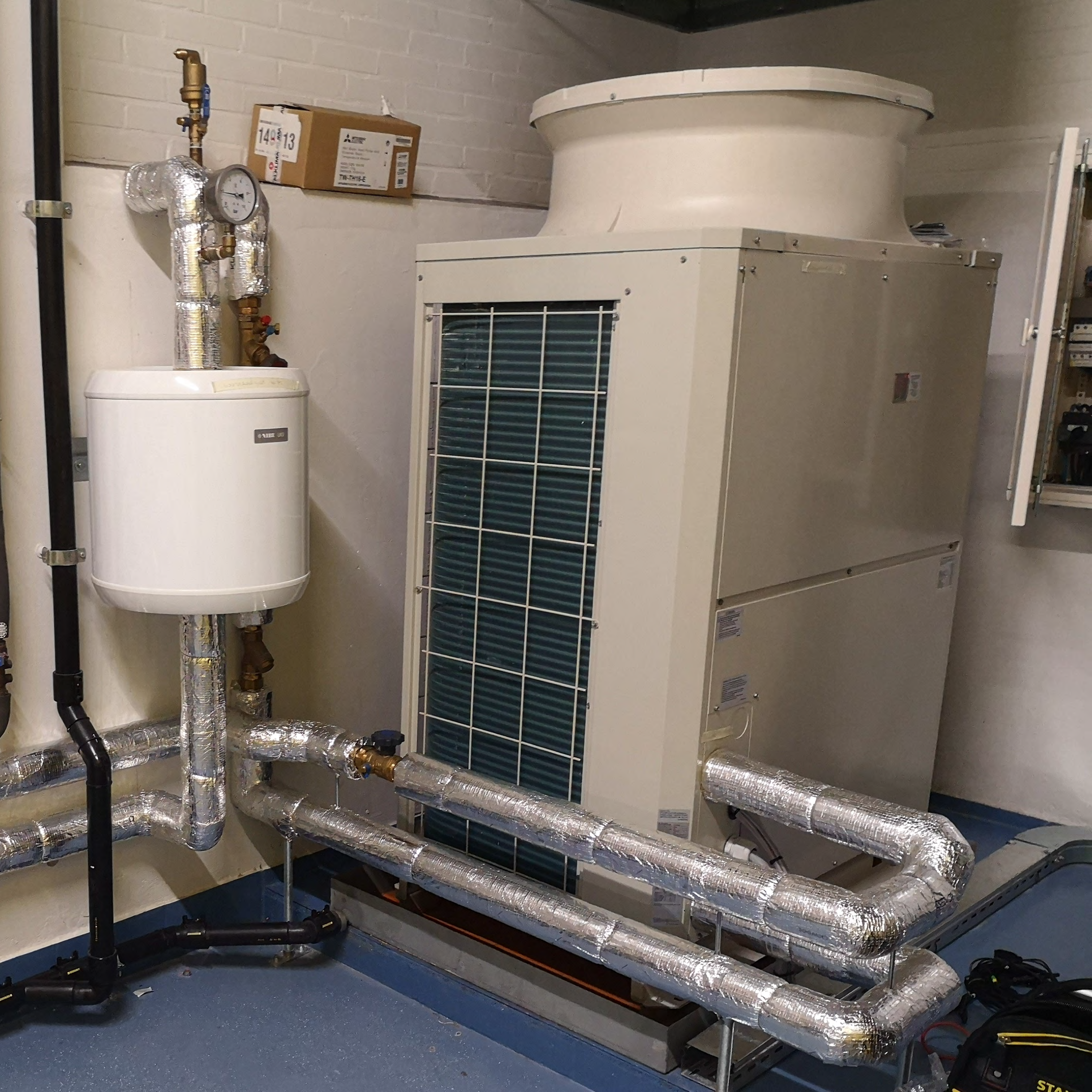}
        \caption{Air-to-water HP.}
        \label{fig:real2}
    \end{subfigure}\quad
    \begin{subfigure}[b]{0.45\linewidth}
    \centering
        \includegraphics[height=0.7\linewidth]{real_system_1.pdf}
        \caption{Two 500L water tanks.}
        \label{fig:real1}
    \end{subfigure}
    \caption{Motivating example of an HPTES system for our numerical case-study.}
    \label{fig:real}
\end{figure}

In order to demonstrate the effectiveness of the proposed scheme, a numerical simulation is performed for an HPTES system, which is based on a real installation for providing domestic hot water in an office building in the Netherlands, see Fig. \ref{fig:real}. This system is composed of one air-to-water HP and two 500L water tanks. The hot water is used for dish washing in dining hall and showering in fitness rooms in the building. The detailed model and corresponding parameters of this HPTES system are provided in Section \ref{sec:appx}.

\begin{figure}[htb]
    \centering
   \includegraphics[width=1\linewidth]{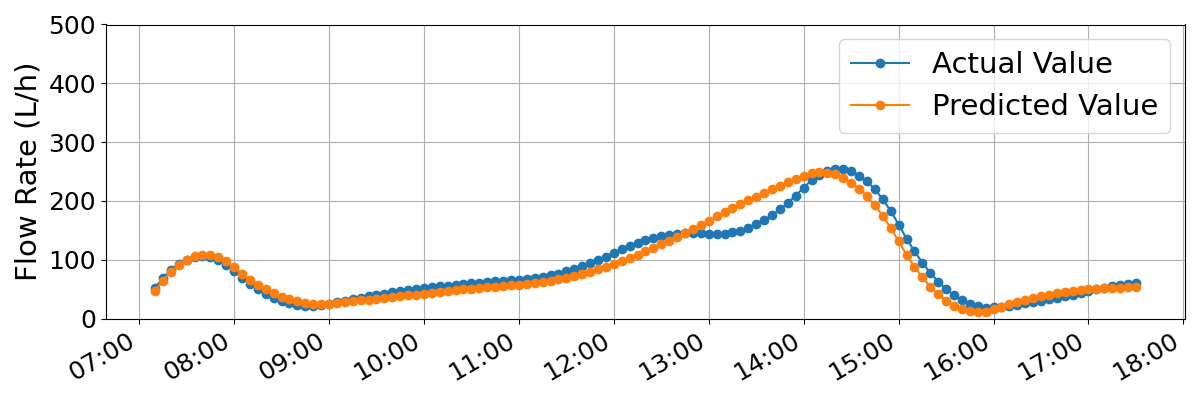}
    \caption{Hot water consumption and its prediction.}
    \label{fig:demand}
\end{figure}

In our case study, the HP is operated during working hours  7:00 am - 5:30 pm. To implement the proposed strategy, hot water consumption $d_t$ and its predicted value $\hat{d}_t$ are needed. Fig. \ref{fig:demand} gives the real hot water consumption profile, which is used for simulating the dynamics of the HPTES system, and the predicted value, which is used for MPC design. The non-negligible prediction errors visible in the figure allow for testing the robustness of our proposed scheme.

To provide qualified hot water, the temperature of the water supplied to buildings, in our case the water temperature at the top of Tank 1 that is denoted as $T_1$, has to satisfy
\begin{equation}\label{eq:T_cons1}
    55 \leq T_1 \leq 75.
\end{equation}
In addition, to mitigate the risk of legionella bacteria, which proliferate between $20^\circ$C - $50^\circ$C, and to provide hot water with sufficient temperature, it is recommended to maintain $T_1$ above $60^\circ$C most of the time:
\begin{equation}\label{eq:T_cons2}
    T_1\geq 60, \text{ for most of the time}.
\end{equation}
Since both constraints \eqref{eq:T_cons1} and \eqref{eq:T_cons2} are hard constraints, to ensure the feasibility of the proposed MPC design, they are relaxed as soft constraints: 
\begin{subequations}\label{eq:soft_cons}
    \begin{align}
        & 55 - \delta_1 \leq T_1  \leq 75+\delta_1, \\
        & T_1 \geq 60 - \delta_2,
    \end{align}
\end{subequations}
where $\delta_1>0$ and $\delta_2>0$ are slack variables that will be penalized in the MPC objection function, see \eqref{eq:dr_comp}.

\begin{figure}[htb]
    \centering
   \includegraphics[width=1\linewidth]{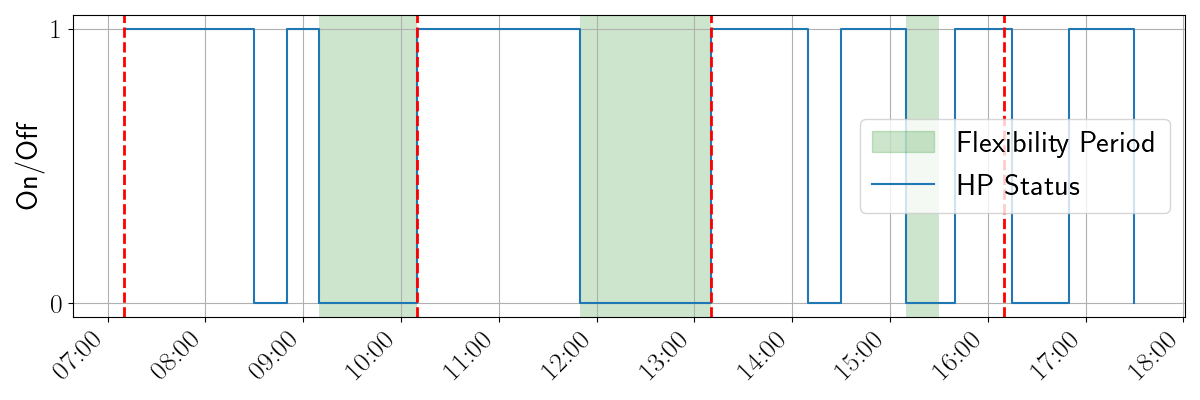}
    \caption{Control input signals of heat pump.}
    \label{fig:action}
\end{figure}

During the whole simulation period, the flexibility task is performed every 3 hours (3 times in total). For flexibility assessment in \eqref{eq:flexi_comp}, the prediction horizon is selected as 4 hours, and the first 3 hours are chosen as flexibility assessment period. The boundaries of flexibility assessment periods are indicated with red dash lines in Fig. \ref{fig:action}. The objective function of the flexibility assessment \eqref{eq:flexi_comp} is chosen as $-\sum_{t\in\mathcal{T}}s_t$, which implies computing the longest period during which HP can remain off without violating system constraints.

\begin{figure}[htb]
    \centering
   \includegraphics[width=1\linewidth]{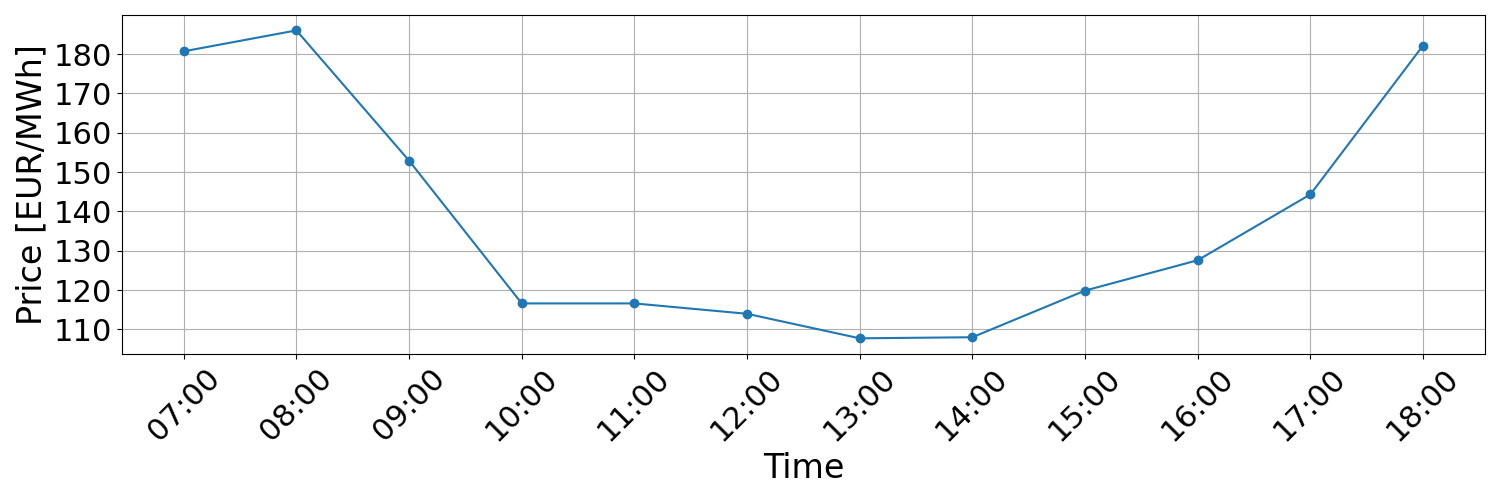}
    \caption{Electricity price used in simulation.}
    \label{fig:price}
\end{figure}

After finishing the flexibility assessment task, the corresponding DR request is set as $\mathcal{R} = \mathcal{F}$. Namely, the power grid tries to utilize all available energy flexibility to reduce electricity consumption during the flexibility period.
For achieving the DR request, the following economic MPC formulation is adopted: 
\begin{subequations}\label{eq:dr_comp}
    \begin{align}
        \min_{u_t,\delta}\ & \sum_{t=0}^N e_tu_t + M_1\delta_1 + M_2\delta_2\\
        \text{s.t. } & x_{t+1} = f(x_{t},u_{t},\hat{d}_{t}),\\
        & x_t \text{ satisfy } \eqref{eq:soft_cons},\ u_t \text{ satisfy }\eqref{eq:switch},\\
        & \forall t = 0,1,2,\cdots, N-1\\
        & u_k = 0, \forall k \in \mathcal{R},
    \end{align}
\end{subequations}
where $e_t$ represents the electricity price at time instant $t$, $\delta_1$ and $\delta_2$ are the slack variables for relaxing the input and state constraints \eqref{eq:soft_cons} if needed, $M_1$ and $M_2$ are corresponding penalties. A typical electricity price in the Netherlands during the simulation period is shown in Fig. \ref{fig:price}, which is used in our simulation. 

\begin{figure}[htb]
    \centering
   \includegraphics[width=1\linewidth]{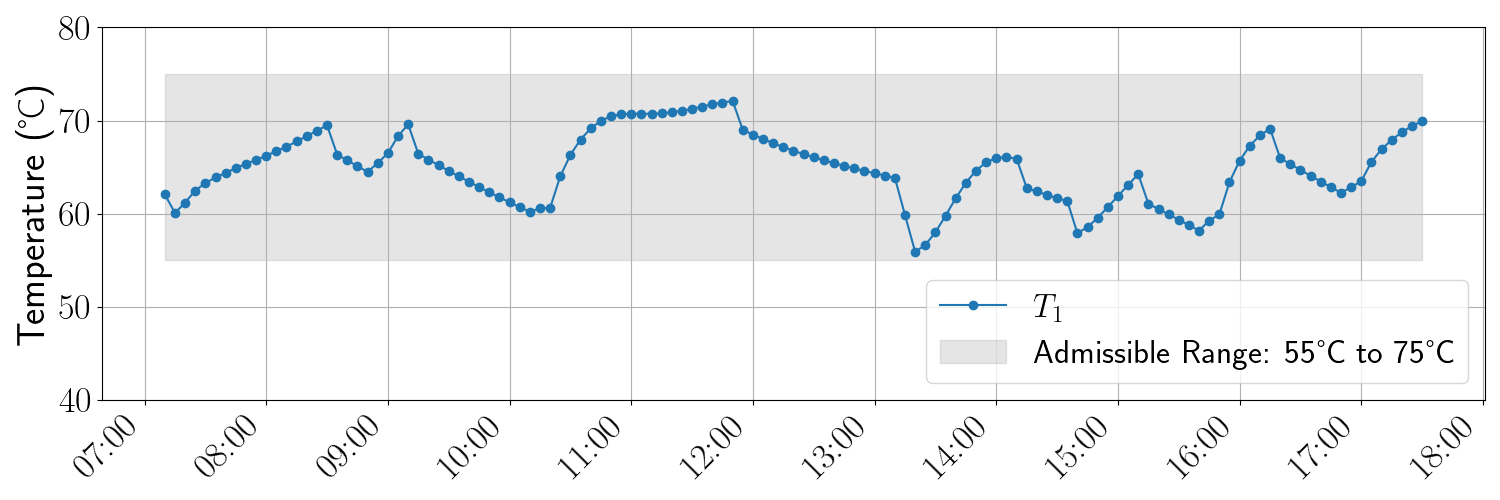}
    \caption{Water temperature at the top of Tank 1.}
    \label{fig:temp}
\end{figure}

Simulation results are given in Fig. \ref{fig:action} - \ref{fig:temp}. Fig. \ref{fig:action} depicts the HP input signals during the simulation period. It can be seen that during all flexibility periods, which are green shaded regions, the HP is kept off to reduce its energy consumption for providing DR services. Fig. \ref{fig:temp} shows the water temperature at the top of Tank 1. It is clear that during the whole simulation period, our proposed approach is able to maintain the temperature within permissable range, which is crucial for providing qualified hot water for buildings. Table \ref{tab:operation_performance} compares the operational performance of the proposed MPC approach with that of the existing rule-based approach implemented for the HPTES installation. With the rule-based approach, the HP is operated only based on the temperature of the water tanks: if water temperature at the bottom of Tank 2 is higher than 62$^\circ$C, HP is off; if temperature at the top of Tank 1 is lower than $62^\circ$C, HP is on. Clearly,
the proposed MPC approach gives better operational performance than the existing rule-based approach even under the presence of flexibility exploitation, which usually degrades the economical performance of MPC. In summary, the proposed MPC approach is not only able to provide DR services by exploiting the energy flexibility of the HPTES system, but is also capable of operating the HPTES system safely and efficiently to provide qualified domestic hot water usage.

\begin{table}[]
    \centering
    \resizebox{\linewidth}{!}{
    \begin{tabular}{lcc}\toprule\hline
         &  Rule-Based Approach & MPC\\\hline
        Average Water Temperature ($^{\circ}$C) & 65.53& 64.85\\
        Maximal Constraint violation ($^{\circ}$C) & 4.43& 4.12\\
         Energy Consumption(kWh) & 64.17 (100\%) & 62.50 (97.40\%)\\
          Energy Cost (Euro) & 8.69 (100\%) & 8.41 (96.78\%)\\\hline
    \end{tabular}}
    \caption{Operational performance of the MPC and the rule-based scheme.}
    \label{tab:operation_performance}
\end{table}

\section{Conclusions}\label{sec:conclusion}
This paper proposed an energy flexible MPC scheme for assessing and exploiting the energy flexibility of HPTES systems to provide demand response services. With the proposed approach, the energy flexibility potential of the HPTES system is quantitatively analyzed by solving a typical mixed-integer economic MPC problem with extra linear constraints. Then the flexibility is activated via feasible DR requests to achieve the reduction of energy consumption without violating system constraints. The proposed scheme provides a systematical approach to take advantage of the flexibility potential of HPTES systems to achieve demand-side management. The efficacy of our design is demonstrated via the simulation results based on a real-world HPTES installation. Future work includes implementing pilot experiments of the proposed scheme on the real-world HPTES installation.

% \section{Acknowledgement}
% The authors 
\bibliographystyle{IEEEtran.bst}
\bibliography{ref}

\begin{thebibliography}{10}
\providecommand{\url}[1]{#1}
\csname url@samestyle\endcsname
\providecommand{\newblock}{\relax}
\providecommand{\bibinfo}[2]{#2}
\providecommand{\BIBentrySTDinterwordspacing}{\spaceskip=0pt\relax}
\providecommand{\BIBentryALTinterwordstretchfactor}{4}
\providecommand{\BIBentryALTinterwordspacing}{\spaceskip=\fontdimen2\font plus
\BIBentryALTinterwordstretchfactor\fontdimen3\font minus \fontdimen4\font\relax}
\providecommand{\BIBforeignlanguage}[2]{{%
\expandafter\ifx\csname l@#1\endcsname\relax
\typeout{** WARNING: IEEEtran.bst: No hyphenation pattern has been}%
\typeout{** loaded for the language `#1'. Using the pattern for}%
\typeout{** the default language instead.}%
\else
\language=\csname l@#1\endcsname
\fi
#2}}
\providecommand{\BIBdecl}{\relax}
\BIBdecl

\bibitem{dutchgov2022}
\BIBentryALTinterwordspacing
D.~Government, ``Klimaatnota 2022,'' 2022. [Online]. Available: \url{https://www.rijksoverheid.nl/documenten/publicaties/2022/11/01/klimaatnota-2022}
\BIBentrySTDinterwordspacing

\bibitem{bunning2022robust}
F.~B{\"u}nning, J.~Warrington, P.~Heer, R.~S. Smith, and J.~Lygeros, ``Robust {MPC} with data-driven demand forecasting for frequency regulation with heat pumps,'' \emph{Control Engineering Practice}, vol. 122, p. 105101, 2022.

\bibitem{Abl:45}
\BIBentryALTinterwordspacing
{European Commission, Directorate General for Energy}, ``Clean energy for all europeans package,'' 2019. [Online]. Available: \url{https://energy.ec.europa.eu/topics/energy-strategy/clean-energy-all-europeans-package_en}
\BIBentrySTDinterwordspacing

\bibitem{hp2022}
\BIBentryALTinterwordspacing
E.~Commission, ``Heat pumps are key to enabling the clean energy transition and achieving the {EU}’s carbon neutrality goal by 2050,'' 2022. [Online]. Available: \url{https://energy.ec.europa.eu/topics/energy-efficiency/heat-pumps_en}
\BIBentrySTDinterwordspacing

\bibitem{ermel2022thermal}
C.~Ermel, M.~V. Bianchi, A.~P. Cardoso, and P.~S. Schneider, ``Thermal storage integrated into air-source heat pumps to leverage building electrification: A systematic literature review,'' \emph{Applied Thermal Engineering}, p. 118975, 2022.

\bibitem{alimohammadisagvand2016cost}
B.~Alimohammadisagvand, J.~Jokisalo, S.~Kilpel{\"a}inen, M.~Ali, and K.~Sir{\'e}n, ``Cost-optimal thermal energy storage system for a residential building with heat pump heating and demand response control,'' \emph{Applied Energy}, vol. 174, pp. 275--287, 2016.

\bibitem{tang2019model}
R.~Tang and S.~Wang, ``Model predictive control for thermal energy storage and thermal comfort optimization of building demand response in smart grids,'' \emph{Applied Energy}, vol. 242, pp. 873--882, 2019.

\bibitem{kircher2015model}
K.~J. Kircher and K.~M. Zhang, ``Model predictive control of thermal storage for demand response,'' in \emph{2015 American Control Conference (ACC)}.\hskip 1em plus 0.5em minus 0.4em\relax IEEE, 2015, pp. 956--961.

\bibitem{renaldi2017optimisation}
R.~Renaldi, A.~Kiprakis, and D.~Friedrich, ``An optimisation framework for thermal energy storage integration in a residential heat pump heating system,'' \emph{Applied energy}, vol. 186, pp. 520--529, 2017.

\bibitem{d2019mapping}
F.~D’Ettorre, M.~De~Rosa, P.~Conti, D.~Testi, and D.~Finn, ``Mapping the energy flexibility potential of single buildings equipped with optimally-controlled heat pump, gas boilers and thermal storage,'' \emph{Sustainable Cities and Society}, vol.~50, p. 101689, 2019.

\bibitem{golmohamadi2022integration}
H.~Golmohamadi, K.~G. Larsen, P.~G. Jensen, and I.~R. Hasrat, ``Integration of flexibility potentials of district heating systems into electricity markets: A review,'' \emph{Renewable and Sustainable Energy Reviews}, vol. 159, p. 112200, 2022.

\bibitem{li2023unlocking}
Y.~Li, N.~Yorke-Smith, and T.~Keviczky, ``Unlocking energy flexibility from thermal inertia of buildings: A robust optimization approach,'' in \emph{2023 62nd IEEE Conference on Decision and Control (CDC)}.\hskip 1em plus 0.5em minus 0.4em\relax IEEE, 2023, pp. 2555--2562.

\bibitem{S22023}
\BIBentryALTinterwordspacing
M.~Konsman, E.~Werkman, and in~collaboration~with TC~205 WG 18~members, ``S2 white paper,'' 2023. [Online]. Available: \url{https://s2standard.org/}
\BIBentrySTDinterwordspacing

\bibitem{rastegarpour2018predictive}
S.~Rastegarpour, M.~Ghaemi, and L.~Ferrarini, ``A predictive control strategy for energy management in buildings with radiant floors and thermal storage,'' in \emph{2018 SICE International Symposium on Control Systems (SICE ISCS)}.\hskip 1em plus 0.5em minus 0.4em\relax IEEE, 2018, pp. 67--73.

\bibitem{junker2018characterizing}
R.~G. Junker, A.~G. Azar, R.~A. Lopes, K.~B. Lindberg, G.~Reynders, R.~Relan, and H.~Madsen, ``Characterizing the energy flexibility of buildings and districts,'' \emph{Applied Energy}, vol. 225, pp. 175--182, 2018.

\bibitem{soren17}
S.~{\O}. Jensen, A.~Marszal-Pomianowska, R.~Lollini, W.~Pasut, A.~Knotzer, P.~Engelmann, A.~Stafford, and G.~Reynders, ``Iea ebc annex 67 energy flexible buildings,'' \emph{Energy and Buildings}, vol. 155, pp. 25--34, 2017.

\end{thebibliography}

\section{Appendix} \label{sec:appx}
In the following, the detailed model and parameters of the HPTES used in our simulation are provided.
For the stratified tank model, Tank 1 and Tank 2 are modeled with 2 and 4 layers, respectively. The thermal dynamics is given in \eqref{eq:sys_model_detail}, the definition of the state variables is given in Table \ref{tab:notation}, and the value of the parameters in \eqref{eq:sys_model_detail} is in Table \ref{tab:parameters}.

\begin{subequations}\label{eq:sys_model_detail}
\begin{align}
    x^1_{k+1} =& u_k \left( x^2_{k+1} + \frac{Q_{\text{hp}}}{m_{\dot{p}} c_p} \right)  + (1 - u_k)\bigg( x^1_k  \nonumber\\
    & - \frac{\Delta t}{m_{\text{pipe}}c_p} \left( R_{\text{pipe}} (x^1_k - T^{\text{amb}}_k) \right.  \nonumber\\
            &\left.  +  R_{\text{pipe,upper}}  (x^1_k - x^3_k) + R_{\text{pipe,bottom}} (x^1_k - x^{8}_k) \right)  \bigg),
            \label{eq:x1_model2}\\
    x^2_{k+1} =& T_{\text{s}} \frac{\dot m_{s}}{\dot m_{p}} + x^{8}_k \left(1 - \frac{\dot m_{s}}{\dot m_{p}}\right),\label{eq:x2_model2}\\
    x^3_{k+1}= & x^3_{k} + \frac{\Delta t}{m_1 c_p} \bigg( \dot{m}_p c_p\left(x^1_k-x^3_k\right) u_k  \nonumber\\
    &-(\dot{m}_c-\dot{m}_s) c_p \triangle T_c   \nonumber\\
    &- R_{12}(x^3_k-x^4_k) + \dot{m}_s c_p(x^4_k-x^3_k)(1-u_k) \bigg)\nonumber \\
&+ \Delta T_{\text{off}}\cdot (u_k-u_{k-1})\cdot u_{k-1} , \label{eq:x3_model2}\\
x^4_{k+1}=&  x^4_k + \frac{\Delta t}{m_2 c_p}\bigg( \left(\dot{m}_p- \dot{m}_s\right) c_p\left(x^3_k-x^4_k\right) u_k  \nonumber\\
&+ R_{12}(x^3_k-x^4_k)-R_{23}(x^4_k-x^5_k)   \nonumber\\
&+ \dot{m}_s c_p(x^5_k-x^4_k)(1-u_k) \bigg) , \label{eq:x4_model2}\\
x^5_{k+1}= & x^7_k + \frac{\Delta t}{m_3 c_p}\bigg( \left(\dot{m}_p- \dot{m}_s\right) c_p\left(x^4_k-x^5_k\right) u_k  \nonumber\\
&+ R_{23}(x^4_k-x^5_k)-R_{34}(x^5_k-x^6_k)  \nonumber\\
&+ \dot{m}_s c_p(x^6_k-x^5_k)(1-u_k)\bigg),\label{eq:x5_model2}\\
x^6_{k+1}= & x^8_k + \frac{\Delta t}{m_4 c_p}\bigg( \left(\dot{m}_p- \dot{m}_s\right) c_p\left(x^5_k-x^6_k\right) u_k  \nonumber\\
&+ R_{34}(x^5_k-x^6_k) - R_{45}(x^6_k-x^7_k)  \nonumber\\
&+ \dot{m}_s c_p(x^7_k-x^6_k)(1-u_k)\bigg), \label{eq:x6_model2}\\
x^7_{k+1}= & x^7_k + \frac{\Delta t}{m_5 c_p}\bigg( \left(\dot{m}_p- \dot{m}_s\right) c_p\left(x^6_k-x^7_k\right) u_k  \nonumber\\
&+ R_{45}(x^6_k-x^7_k) - R_{56}(x^7_k-x^{8}_k)  \nonumber\\
&+ \dot{m}_s c_p(x^{8}_k-x^7_k)(1-u_k)\bigg), \label{eq:x7_model2}\\
x^{8}_{k+1}=&  x^{8}_k + \frac{\Delta t}{m_6 c_p}\bigg( \left(\dot{m}_p- \dot{m}_s\right) c_p\left(x^7_k-x^{8}_k\right) u_k  \nonumber\\
&+ R_{56}(x^7_k-x^{8}_k) + \dot{m}_s c_p(T_s-x^{8}_k)(1-u_k)\bigg), \\
Q_{hp} =&  \text{COP} \cdot P_r \cdot u, \\
\text{COP} = & a_1 + a_2T_{\text{in,hp}} + a_3 T_{\text{amb}} + a_4T_{\text{in,hp}}T_{\text{amb}},\\
T_{\text{in,hp}} = & \Delta T_{\text{he}} + T_{\text{out,tank}}.\label{eq:14k}
\end{align}
\end{subequations}
In \eqref{eq:sys_model_detail}, the thermal loss $\dot{Q}_{\text{wall},j}$ in \eqref{eq:tank_model} is omitted since the HPTES system is well-insulated such that the thermal loss is negligible. Equality \eqref{eq:14k} is derived via 
\begin{equation}
        T_{\text{out,hp}} - T_{\text{in,hp}} = T_{\text{in,tank}} - T_{\text{out,tank}}
\end{equation}
with $\Delta T_{\text{he}}:= T_{\text{out,hp}} - T_{\text{in,tank}}$ as a relatively stable temperature difference, which is true when the heat exchanger is operated with high efficiency.

In addition, the term $\Delta T_{\text{off}}\cdot(u_k - u_{k-1})\cdot u_{k-1}$ in \eqref{eq:x3_model2} is used to reflect a special phenomenon of the temperature drop $\Delta T_{\text{off}}$ for water in the top of Tank 1 when HP is switched from on to off, which is caused by the HP and the circulation pump of water tanks being out of sync in our HPTES system.

% {\color{blue}
% Among them, according to \eqref{eq:hp} and \eqref{eq:cop2}, $Q_{hp}$ can be related to $T_{\text{in,hp}}$ and $T_{\text{amb}}$. $T_{\text{amb}}$ can be obtained through a temperature sensor. $T_{\text{in,hp}}$ is obtained through the relationship with the heat exchanger, via $T_{\text{out,tank}}$

% According to the manufacturer’s documentation of the heat pump system, effective heat exchange is achieved under optimal conditions with a 5°C temperature difference $\Delta T_{\text{he}}$ between the heat pump’s outlet water temperature $T_{\text{out,hp}}$ and the water tank’s inlet temperature $T_{\text{in,tank}}$. Furthermore, under ideal conditions, the temperature difference between the heat pump’s outlet and inlet water temperatures, and the temperature difference between the water tank’s inlet and outlet water temperatures are the same. Thus, we have the formula:
% \begin{equation}
%     \Delta T_{\text{he}} = T_{\text{out,hp}} - T_{\text{in,tank}}
% \end{equation}
% \begin{equation}
%     T_{\text{out,hp}} - T_{\text{in,hp}} = T_{\text{in,tank}} - T_{\text{out,tank}}
% \end{equation}
% Therefore, we can calculate $T_{\text{in,hp}}$ through $T_{\text{out,tank}}$ as follows:
% \begin{equation}
%     \begin{aligned}
%         T_{\text{in,hp}} &= (T_{\text{out,hp}} - T_{\text{in,tank}}) + T_{\text{out,tank}}\\
%         & = \Delta T_{\text{he}}+ T_{\text{out,tank}}
%     \end{aligned}
% \end{equation}
% }

\begin{table}[h!]
    \centering
    \begin{tabular}{llc}\toprule\hline
    State& Description & Units \\\hline
$x^1$ & \begin{tabular}{l} 
The inlet water temperature of\\
the water tanks, $T_{\text{in,tank}}$
\end{tabular}  & $^{\circ}$C \\
 $x^2$ & \begin{tabular}{l} 
The outlet water temperature of\\
the water tanks $T_{\text{out,tank}}$
\end{tabular}  & $^{\circ}$C \\
 $x^3$ & \begin{tabular}{l} 
The temperature of the 1st layer of\\
the water tanks $T_1$
\end{tabular}  & $^{\circ}$C \\
 $x^4$ & \begin{tabular}{l} 
The temperature of the 2nd layer of\\
the water tanks $T_2$
\end{tabular}  & $^{\circ}$C \\
 $x^5$ & \begin{tabular}{l} 
The temperature of the 3rd layer of\\
the water tanks $T_3$
\end{tabular} & $^{\circ}$C \\
  $x^6$ & \begin{tabular}{l} 
The temperature of the 4th layer of\\
the water tanks $T_4$
\end{tabular}  & $^{\circ}$C \\
  $x^7$ & \begin{tabular}{l} 
The temperature of the 5th layer of\\
the water tanks $T_5$
\end{tabular}  & $^{\circ}$C \\
  $x^{8}$ & \begin{tabular}{l} 
The temperature of the 6th layer of\\
the water tanks $T_6$
\end{tabular}  & $^{\circ}$C \\
\hline
\end{tabular}
    \caption{Description of notations in \eqref{eq:sys_model_detail}.}
    \label{tab:notation}
\end{table}
% \begin{table}[h!]
%     \centering
%     \begin{tabular}{llc}\toprule\hline
%     State& Description & Units \\\hline
% $x^1$ & \begin{tabular}{l}The outlet water temperature of \\
% the heat pump $T_{out,hp}$\end{tabular} & $^{\circ}$C \\
%  $x^2$ & \begin{tabular}{l} 
% The inlet water temperature of \\
% the heat pump $T_{in,hp}$
% \end{tabular}  & $^{\circ}$C \\
%  $x^3$ & \begin{tabular}{l} 
% The inlet water temperature of\\
% the tank $T_{in,tank}$
% \end{tabular}  & $^{\circ}$C \\
%  $x^4$ & \begin{tabular}{l} 
% The outlet water temperature of\\
% the tank $T_{out,tank}$
% \end{tabular}  & $^{\circ}$C \\
%  $x^5$ & \begin{tabular}{l} 
% The temperature of the 1st layer of\\
% the water tanks $T_1$
% \end{tabular}  & $^{\circ}$C \\
%  $x^6$ & \begin{tabular}{l} 
% The temperature of the 2nd layer of\\
% the water tanks $T_2$
% \end{tabular}  & $^{\circ}$C \\
%  $x^7$ & \begin{tabular}{l} 
% The temperature of the 3rd layer of\\
% the water tanks $T_3$
% \end{tabular} & $^{\circ}$C \\
%   $x^8$ & \begin{tabular}{l} 
% The temperature of the 4th layer of\\
% the water tanks $T_4$
% \end{tabular}  & $^{\circ}$C \\
%   $x^9$ & \begin{tabular}{l} 
% The temperature of the 5th layer of\\
% the water tanks $T_5$
% \end{tabular}  & $^{\circ}$C \\
%   $x^{10}$ & \begin{tabular}{l} 
% The temperature of the 6th layer of\\
% the water tanks $T_6$
% \end{tabular}  & $^{\circ}$C \\
% \hline
% \end{tabular}
%     \caption{Description of notations in \eqref{eq:sys_model_detail}.}
%     \label{tab:notation}
% \end{table}

\begin{table}[h!]
    \centering
    \resizebox{0.48\textwidth}{!}{
    \begin{tabular}{llcc}\toprule
\hline Term & Description & Value & Units \\
\hline $\Delta t$ & sampling time & 1/3 & h \\
\hline$c_p$ & \begin{tabular}{l} 
Specific heat capacity of water
\end{tabular} & 4186 & $\mathrm{J/kg\cdot K}$  \\
\hline$R_{\text{pipe}}$ & \begin{tabular}{l} 
Thermal resistance for outlet pipe \\ 
of the heat exchanger
\end{tabular} & 0.30 & $\mathrm{W/K}$ \\
\hline$R_{\text{pipe,upper}}$ & \begin{tabular}{l} 
Thermal resistance between \\
outlet pipes of the heat exchanger \\
and the upper layer of Tank 1
\end{tabular} & 0 & $\mathrm{W/K}$ \\
\hline$R_{\text{pipe,bottom}}$ & \begin{tabular}{l} 
Thermal resistance between  \\
outlet pipes of the heat exchanger  \\
and the bottom layer of Tank 2
\end{tabular} & 2 & $\mathrm{W/K}$ \\
\hline$R_{12}$ & \begin{tabular}{l} 
Thermal resistance between the 1st\\
 and 2nd layers of the water tanks
\end{tabular} & 0.24 & $\mathrm{W/K}$ \\
\hline$R_{23}$ & \begin{tabular}{l} 
Thermal resistance between the 2nd\\
 and 3rd layers of the water tanks
\end{tabular} & 0.24 & $\mathrm{W/K}$ \\
\hline$R_{34}$ & \begin{tabular}{l} 
Thermal resistance between the 3rd\\
 and 4th layers of the water tanks
\end{tabular} & 0.49 & $\mathrm{W/K}$ \\
\hline$R_{45}$ & \begin{tabular}{l} 
Thermal resistance between the 4th\\
 and 5th layers of the water tanks
\end{tabular} & 0.54 & $\mathrm{W/K}$ \\
\hline$R_{56}$ & \begin{tabular}{l} 
Thermal resistance between the 5th\\
and 6th layers of the water tanks
\end{tabular} & 0.53 & $\mathrm{W/K}$ \\
\hline$m_{\text{pipe}}$ & \begin{tabular}{l} 
Mass of the outlet pipe of \\ the heat exchanger 
\end{tabular} & 3.27 & kg \\
\hline$m_{1}$ & Mass of the 1st layers of the water tanks & 250 & kg \\
\hline$m_{2}$ & Mass of the 2nd layers of the water tanks & 250 & kg \\
\hline$m_{3}$ & Mass of the 3rd layers of the water tanks & 169.66 & kg \\
\hline$m_{4}$ & Mass of the 4th layers of the water tanks & 95.38 & kg \\
\hline$m_{5}$ & Mass of the 5th layers of the water tanks & 136.67 & kg \\
\hline$m_{6}$ & Mass of the 6th layers of the water tanks & 98.29 & kg \\
\hline$\Delta T_{\text{he}}$ & Temperature difference of the heat exchanger & 2.84 & $^\circ$C \\
\hline$\Delta T_{c}$ & Temperature difference in the demand pipe & 1.76 & $^\circ$C \\
\hline$T_s$ & Temperature of the water in supply pipe & 13 & $^\circ$C \\
\hline$\Delta T_{\text{off}}$ & \begin{tabular}{l} 
Temperature drop of \\
the heat pump's deactivation
\end{tabular} & 2.5 & $^\circ$C \\
\hline $T_{\text{amb}}$ & Ambient Temperature & 18.5 &  $^\circ$C\\
\hline$ \dot m_{s}$ & Flow rate in the supplement pipe & - & kg/h \\
\hline$ \dot m_{c}$ & Flow rate in the demand circulation pipe & 1100 & $\mathrm{kg/h}$\\
\hline$ \dot{m}_{p}$ & Flow rate in the main circulation pipe & 880 & $\mathrm{kg/h}$\\
\hline$a_1$ & Parameter in COP & 3.3297 & - \\
\hline$a_2$ & Parameter in COP & -0.0423 & -\\
\hline$a_3$ & Parameter in COP & 0.0219 & -\\
\hline$a_4$ & Parameter in COP & 0.0003 & - \\
\hline
\end{tabular}}
    \caption{Nominal values of the parameters in \eqref{eq:sys_model_detail}.}
    \label{tab:parameters}
\end{table}

\end{document}